\documentclass[preprint,12pt]{elsarticle}

\usepackage{amssymb}

\usepackage{amsmath}
\usepackage{float}

\journal{arxiv}

\DeclareMathOperator{\Tr}{Tr}

\begin{document}

\begin{frontmatter}



\title{On the validity of intermediate tracing in multiple quantum interactions}


\author[inst1,inst4]{Reuven Ianconescu}
\author[inst2,inst4]{Bin Zhang}
\author[inst3]{Aharon Friedman}
\author[inst4,inst5]{Jacob Scheuer}
\author[inst4,inst5]{Avraham Gover}

\affiliation[inst1]{organization={Shenkar College of Engineering and Design, Department of Electrical Engineering},
            addressline={Anna Frank 12}, 
            city={Ramat Gan},
            postcode={5252626},
            state={Israel}
}

\affiliation[inst2]{organization={State Key Laboratory of Quantum Functional Materials, School of Physical Science and Technology and Center for Transformative Science, ShanghaiTech University},
            addressline={Anna Frank 12}, 
            city={Shanghai},
            postcode={200031},
            state={China}
}

\affiliation[inst3]{organization={Schlesinger Family Accelerator Centre, Ariel University},
            addressline={Anna Frank 12}, 
            city={Ariel},
            state={Israel}
}

\affiliation[inst4]{organization={School of Electrical Engineering–Physical Electronics, Tel Aviv University},
            addressline={Anna Frank 12}, 
            city={Ramat Aviv, Tel Aviv},
            state={Israel}
}

\affiliation[inst5]{organization={Center of Laser-Matter Interaction, Tel Aviv University},
            addressline={Anna Frank 12}, 
            city={Ramat Aviv, Tel Aviv},
            state={Israel}
}



\begin{abstract}
Interactions between many (initially separate) quantum systems raise the question on how to prepare and how to compute the measurable results of their interaction. When one prepares each system individually and let them interact, one has to tensor multiply their density matrices and apply Hamiltonians on the composite system (i.e. the system which includes all the interacting systems) for definite time intervals. Evaluating the final state of one of the systems after multiple consecutive interactions, requires tracing all other systems out of the composite system, which may grow up to immense dimensions. For computation efficiency during the interaction(s) one may consider only the contemporary interacting partial systems, while tracing out the other non interacting systems. In concrete terms, the type of problems to which we direct this formulation is a ``target'' system interacting {\bf succesively} with ``incident'' systems, where the ``incident'' systems do not mutually interact. For example a two-level atom, interacting succesively with free electrons, or a resonant cavity interacting with radiatively free electrons, or a quantum dot interacting succesively with photons. We refer to a ``system'' as one of the components before interaction, while each interaction creates a ``composite system''. A new interaction of the ``composite system'' with another ``system'' creates a ``larger composite system'', unless we trace out one of the systems before this interaction.  The scope of this work is to show that under proper conditions one may add a system to the composite system just before it interacts, and one may trace out this very system after it finishes to interact. We show in this work a mathematical proof of the above property and give a computational example.
\end{abstract}




\end{frontmatter}


\section{Introduction}

Quantum interactions between multiple systems may require a big amount of computing resources, depending on the number of systems and their dimensionality. The question at which stage to add a system to the composite system and at which state to trace it out is relevant.

Tracing degrees of freedom is common practice in the domain of multi-system interaction. This procedure has been used in optical excitations with electron beams \citep{optical_exc}, where it is shown that the optical excitation probability by a single electron is independent of its wave function, while the probability for more (modulated) electrons depends on their relative spatial arrangement, thus reflecting the quantum nature of their interactions. Entanglement between photons induced by free electrons is analyzed in \citep{photon_ent}, showing that free electrons can control the second-order coherence of initially independent photonic states, even in spatially separated cavities that cannot directly interact. In \citep{ultrafast} it is shown how precise control of the electron before and after its interaction with quantum light enables to extract the photon statistics and implement full quantum state tomography using PINEM (Photon-Induced Near-field Electron Microscopy).

In our works we dealt with such multi-system interactions, as follows. In \citep{duality} we analyzed the interaction between two quantum systems, one free electron and one bound electron modeled as a TLS (Two level system). This type of interaction has been labeled FEBERI (Free-Electron Bound Electron Resonant Interaction), as elaborated in \citep{feberi}. The analysis delineated the particle-like and wave-like interaction regimes and discussed the possibility of using the free electron wavepacket for interrogation and coherent control of the TLS. In \citep{coherent} we analyzed the coherent excitation of a TLS by multiple free electrons modeled as quantum wave packets. To learn the accumulated effect on the TLS we each time traced out the free electrons. We found that the transition probability of the TLS grows quadratically with the number of correlated quantum electron wavepackets (which correspond to the quadratic expansion of the sinusoidal transition rate in a Rabi oscillation process). In \citep{interrog} we used the above formalism to interrogate on the state of a TLS using preshaped free electron quantum wavepackets. Measurement of the post-interaction energy spectrum of the free electrons probes and quantifies the full Bloch sphere parameters of the TLS. Interesting studies in electron-induced excitation of whispering gallery modes have been presented in \citep{auad, muller, kfir1, kfir2, kfir3,karnieli}. In \citep{spont_bin} we used the same formalism to examine the spontaneous emission of photons by pre-shaped quantum wave packets, analyzed the relation between the photon's density matrix (or its Wigner distribution representation) and the quantum electrons wavefunctions, which revealed the quantum process origin of the evolution of bunched beam superradiance \citep{rmp}. The reality of the quantum electron wave packet (QEW) and the measurability of its dimensions, as well as the transition from the quantum wave function presentation to the classical point-particle theory (the wave-particle duality) were considered previously in the context of electron interactions with light \citep{friedman,pan,fares,corson,kaminer,remez}.

This work is theoretical, and is applicable to theoretical and computational studies. We carry out an analytic calculation which proves in generality that for the purpose of examining the results of quantum interactions between multiple systems it is enough to add a system to the composite system before it interacts, and it is valid to trace it out after it finishes its interaction. As explained in the abstract, we consider the interaction between separate systems, meaning they are initially separable. During interaction they usually become entangled and hence are not separable any more, meaning that the measurement probabilities on those systems are correlated. To find out the changes on each system, we trace out the other systems and analyze each one separately, as discussed in section~3.4 . In section~2 we carry out the analytic proof and in section~3 we present a numerical example with TLSs to show how this works for the simplest case of three consecutively interacting systems. The work is ended by some concluding remarks.
\section{Analytic proof}

We want to examine at which stage we have to add a new system to the composite system and at which stage we may trace it out. Certainly a system has to be present in the composite system at least during its interaction, but here we show that it has to be in the composite system {\bf only} during its interaction. For this purpose it is enough to consider two systems. The first (named below A) represents the ``target'' system interacting with one of the ``incident'' systems and the second (named below B) is another ``incident'' system. We consider the density matrix of the ``incident'' systems to be known before the interaction, therefore we do not evolve here system B. Therefore, the interaction happens inside system A {\bf only}, and we show that its results do not depend whether the ``other'' system (B) has been added to it.

We first consider system A alone, described by the density matrix $\rho_A$, dynamically changing according to the Hamiltonian $H_A$. The evolution of A is described by the equation of motion
\begin{equation}
\frac{d\rho_A}{dt} = i(\rho_A H_A-H_A\rho_A)
\label{eq_motion_A}
\end{equation}
If system B, described by the density matrix $\rho_B$ has been added to A, the composite system is described by
\begin{equation}
\rho_S=\rho_A\otimes\rho_B,
\label{rho}
\end{equation}
so that the individual systems density matrices can be obtained by tracing over the coordinates of the ``other'' system
\begin{equation}
\rho_A=\Tr_B\rho_S
\label{rho_A}
\end{equation}
and
\begin{equation}
\rho_B=\Tr_A\rho_S.
\label{rho_B}
\end{equation}
As shown in Eq.~(\ref{eq_motion_A}), the interactions inside system A are governed by the Hamiltonian $H_A$. Knowing that system B does not interact with A, we could use any Hamiltonian of the type $H_A\otimes U+U\otimes H_B$, but as explained before we are not interested in the evolution of B, therefore we use the following Hamiltonian
\begin{equation}
H_S=H_A\otimes U,
\label{H}
\end{equation}
where $U$ is the unit operator.

The equation of motion of the system S is
\begin{equation}
\frac{d\rho_S}{dt}=i[\rho_S,H_S]=i(\rho_S H_S-H_S\rho_S)
\label{eq_motion}
\end{equation}
Using Eq.~(\ref{rho}), the LHS of (\ref{eq_motion}) is
\begin{equation}
\frac{d\rho_S}{dt}=\frac{d}{dt}(\rho_A\otimes\rho_B)=\frac{d\rho_A}{dt}\otimes\rho_B+\rho_A\otimes\frac{d\rho_B}{dt}
\label{eq_motion_LHS}
\end{equation}
Using Eq.~(\ref{H}) and the mixed-product property $(A \otimes B)(C \otimes D) = (AC) \otimes (BD)$, the RHS of Eq.~(\ref{eq_motion}) consists of
\begin{equation}
\rho_S H_S=(\rho_A\otimes\rho_B)(H_A\otimes U)=(\rho_A H_A)\otimes(\rho_BU)=(\rho_A H_A)\otimes \rho_B
\label{RHS_1}
\end{equation}
\begin{equation}
H_S\rho_S=(H_A\otimes U)(\rho_A\otimes\rho_B)=(H_A\rho_A)\otimes(U\rho_B)=(H_A\rho_A)\otimes \rho_B
\label{RHS_2}
\end{equation}
Putting those together, we obtain the equation of motion
\begin{equation}
\frac{d\rho_A}{dt}\otimes\rho_B+\rho_A\otimes\frac{d\rho_B}{dt}=i(\rho_A H_A-H_A\rho_A)\otimes \rho_B.
\label{eq_motion_1}
\end{equation}
Tracing Eq.~(\ref{eq_motion_1}) over the coordinates of B, using the properties $\Tr\rho=1$, and therefore $\Tr\frac{d\rho}{dt}=\frac{d}{dt}\Tr\rho=0$, we recover Eq.~(\ref{eq_motion_A}), showing that the evolution of A is not affected by the presence of B. Tracing Eq.~(\ref{eq_motion_1}) over the coordinates of A, results in
\begin{equation}
\frac{d\rho_B}{dt}=0,
\label{eq_motion_B}
\end{equation}
showing that system B is not affected. In the following section we show on a simple example how this principle works.

\section{Numerical example}

We emphasize that the ideas presented in this work are general and applicable to a large series of problems, as explained in the Abstract and shown in the Introduction. We present here a simple example (out of many possible examples) to demonstrate the usage of those ideas.

For this example we use 3 qubits: A, B and C, in a model of pure spin-spin interaction between pairs. First we interact qubits A and B, and after this interaction finishes we interact qubits A and C. The Hamiltonian for the spin-spin interaction is $H=\boldsymbol\sigma\cdot\boldsymbol\sigma$, where the scalar multiplication implies the sum of the multiplications of all the spin components.

This relates to the proof in the previous section as follows. Say we interact qubits A and B, so we may call this interacting system AB (named in section~2 A). Qubit C does not interact here, so this is the "other" system (named in section~2 B). Either qubit C is part of the system as in Policy 1 or outside it as in Policy 3, the results come out identical.

\begin{figure}[H]
\includegraphics[width=13 cm]{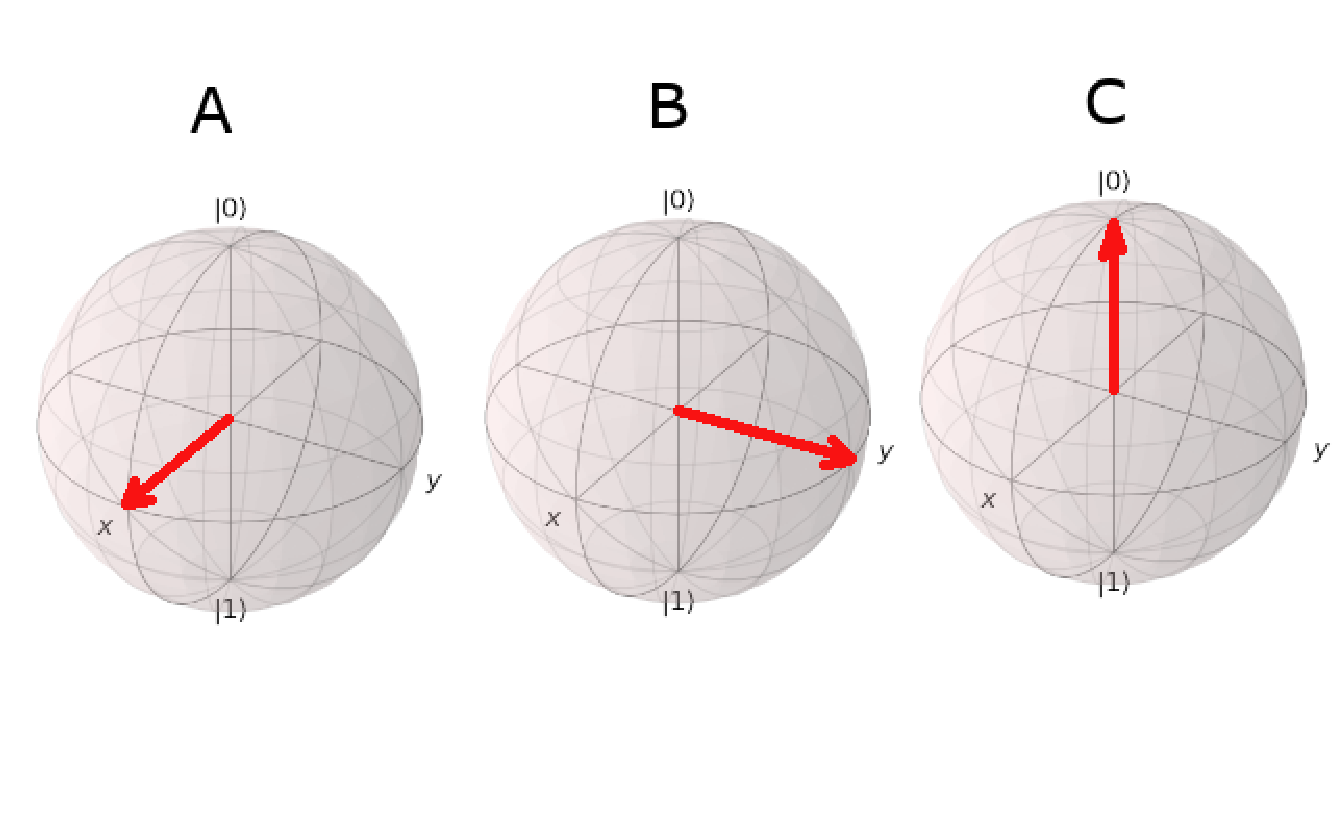}
\caption{The initial configuration of the qubits A, B and C: in the positive eigenstate of $\sigma_x$, $\sigma_y$ and $\sigma_z$ respectively.}
\label{configuration}
\end{figure}

The initial configuration of the qubits is shown in Figure~\ref{configuration} as Bloch spheres. Qubits A, B and C are in the positive eigenstate of $\sigma_x$, $\sigma_y$ and $\sigma_z$ respectively.
To run an interaction we implement the equation of motion using the following recursion
\begin{equation}
\rho^{(n+1)}=\rho^{(n)}+i\, dt\,\left(\rho^{(n)} H-H\rho^{(n)}\right)
\label{eq_motion_numeric}
\end{equation}
for 500 steps, each step advances the time by $dt=1\times 10^{-4}$ (in natural units $\hbar=1$).
We run the interactions using 3 different policies of adding or tracing out qubits, as shown in the subsections below.

\subsection{Policy 1}

We show here the least efficient policy, to keep all components in the system. We combine the whole system of 3 qubits
\begin{equation}
\rho_{ABC}=\rho_A\otimes\rho_B\otimes\rho_C
\label{build_system1}
\end{equation}
Then we interact A with B, using
\begin{equation}
H=\boldsymbol{\sigma}\cdot\boldsymbol{\sigma}\cdot\mathbf{1}=\sigma_x\otimes\sigma_x\otimes U+\sigma_y\otimes\sigma_y\otimes U+\sigma_z\otimes\sigma_z\otimes U
\label{H_AB_1}
\end{equation}
After this interaction finished, we interact A and C, using
\begin{equation}
H=\boldsymbol{\sigma}\cdot\mathbf{1}\cdot\boldsymbol{\sigma}=\sigma_x\otimes U\otimes \sigma_x+\sigma_y\otimes U\otimes \sigma_y+\sigma_z\otimes U\otimes \sigma_z
\label{H_AC}
\end{equation}
At the end we partial trace
\begin{equation}
\rho_A=\Tr_{BC}\{\rho_{ABC}\}\,\,\,\,\,;\,\,\,\,\,\rho_B=\Tr_{AC}\{\rho_{ABC}\}\,\,\,\,\,;\,\,\,\,\,\rho_C=\Tr_{AB}\{\rho_{ABC}\}
\label{trace_1}
\end{equation}

\subsection{Policy 2}
This policy has a better efficiency than the previous one, but is not the best possible. Like in the previous policy, we build the whole system
\begin{equation}
\rho_{ABC}=\rho_A\otimes\rho_B\otimes\rho_C
\label{build_system2}
\end{equation}
and interact A with B, using
\begin{equation}
H=\boldsymbol{\sigma}\cdot\boldsymbol{\sigma}\cdot\mathbf{1}=\sigma_x\otimes\sigma_x\otimes U+\sigma_y\otimes\sigma_y\otimes U+\sigma_z\otimes\sigma_z\otimes U.
\label{H_AB_2}
\end{equation}
Unlike the previous case, after this interaction finished, we trace out B, obtaining the density matrix of system AC
\begin{equation}
\rho_{AC}=\Tr_{B}\{\rho_{ABC}\}
\label{rho_AC}
\end{equation}
and trace out AC to obtain the density matrix of B
\begin{equation}
\rho_{B}=\Tr_{AC}\{\rho_{ABC}\}
\end{equation}
The advantage of this policy vs the previous one is in the interaction that follows, we use only 2 qubits instead of 3. So we interact the system of 2 qubits AC, using
\begin{equation}
H=\boldsymbol{\sigma}\cdot\boldsymbol{\sigma}=\sigma_x\otimes \sigma_x+\sigma_y\otimes \sigma_y+\sigma_z\otimes \sigma_z
\label{H_AC_2}
\end{equation}

At the end we partial trace to obtain the density matrices of A and C
\begin{equation}
\rho_A=\Tr_{C}\{\rho_{AC}\}\,\,\,\,\,;\,\,\,\,\,\rho_C=\Tr_{A}\{\rho_{AC}\}
\label{rho_A_2}
\end{equation}

\subsection{Policy 3}
This is the most efficient policy, we keep each time only the interacting components. First we build the partial system AB
\begin{equation}
\rho_{AB}=\rho_A\otimes\rho_B,
\label{rho_AB_3}
\end{equation}
and interact A with B, using
\begin{equation}
H=\boldsymbol{\sigma}\cdot\boldsymbol{\sigma}=\sigma_x\otimes\sigma_x+\sigma_y\otimes\sigma_y+\sigma_z\otimes\sigma_z
\label{H_AB_3}
\end{equation}
After interaction finished, we trace out B, remaining with
\begin{equation}
\rho_{A}=\Tr_{B}\{\rho_{AB}\}
\label{rho_A_3}
\end{equation}
and trace out A obtaining
\begin{equation}
\rho_{B}=\Tr_{A}\{\rho_{AB}\}.
\label{rho_B_3}
\end{equation}
Now we build the system
\begin{equation}
\rho_{AC}=\rho_A\otimes\rho_C,
\label{rho_AC_3}
\end{equation}
and interact A with C, using the above Hamiltonian. At the end we partial trace to obtain the density matrices for A and C
\begin{equation}
\rho_A=\Tr_{C}\{\rho_{AC}\}\,\,\,\,\,;\,\,\,\,\,\rho_C=\Tr_{A}\{\rho_{AC}\}
\label{rho_A_rho_C_3}
\end{equation}

As the theory predicts, all three policies give identical results for the 3 qubits, here are the results in qubit parameters radius ($r$), elevation angle in degrees ($\theta$) and azimuth angle in degrees($\varphi$).

Qubit A
\begin{equation}
r=0.98913\,\,\,\,\,\,\theta=95.072\,\,\,\,\,\varphi=6.3053
\label{qubit_A}
\end{equation}

Qubit B
\begin{equation}
r=0.99507\,\,\,\,\,\,\theta=84.299\,\,\,\,\,\varphi=89.424
\label{qubit_B}
\end{equation}

Qubit C
\begin{equation}
r=0.99399\,\,\,\,\,\,\theta=8.481\,\,\,\,\,\varphi=-83.706
\label{qubit_C}
\end{equation}

\begin{figure}[H]
\includegraphics[width=13 cm]{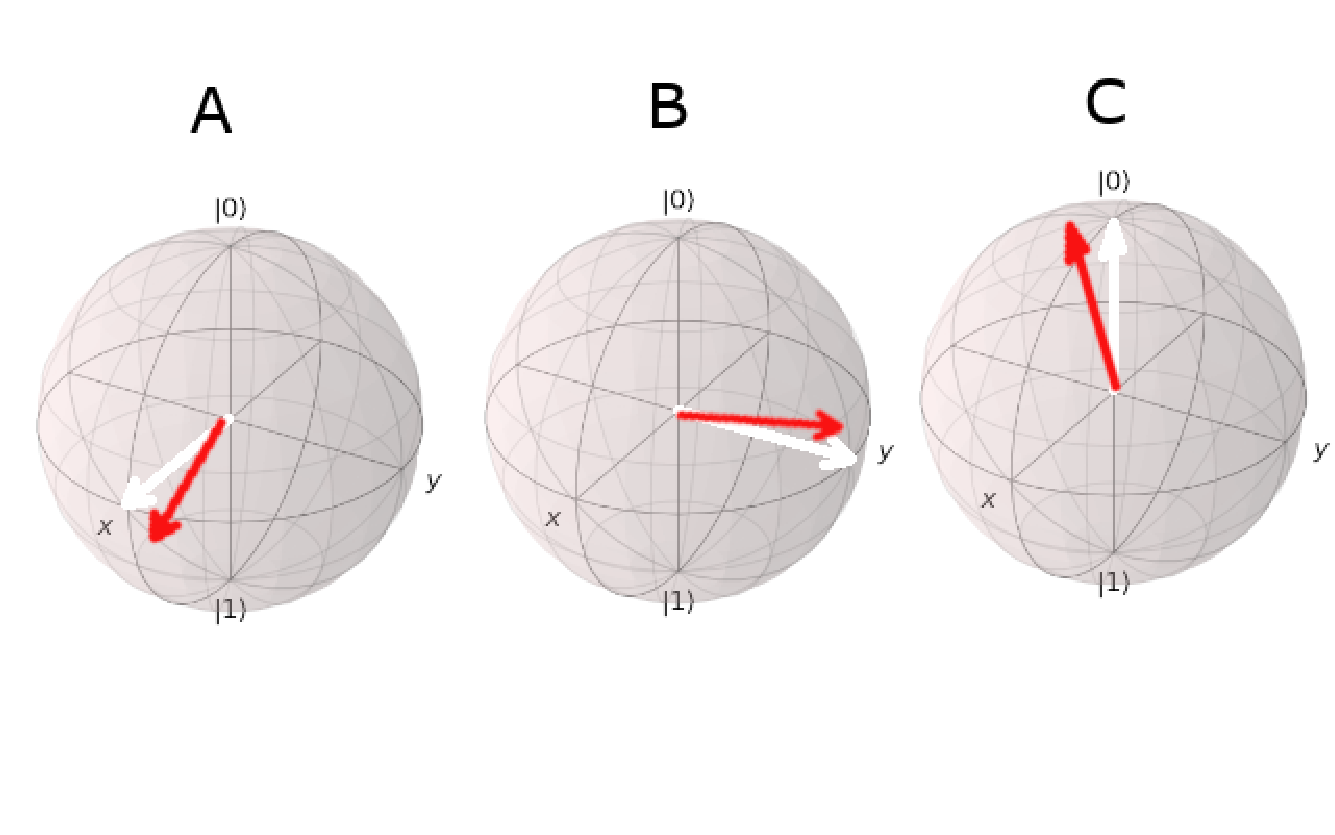}
\caption{The 3 qubits after the completion of all interactions, their states are marked by the red arrows. For reference, the white arrows show the states of the qubits before interaction.}
\label{result}
\end{figure}

The comparison between the run time of those algorithms shows that policy 2 is by 30\% more effective than policy 1, and policy 3 is by 60\% more effective than policy 1, which makes sense because in policy 2 we made one calculation more efficient, while in policy 3 we made the calculations for both interactions more efficient.
The results are show in Figure~\ref{result}, the red arrows show the state of the qubits after interaction, and the white arrows (for reference) show the state of the qubits before interaction.

\subsection{Analysis of the results}

We interpret here the results of the above interactions. The probability to measure a qubit in the positive eigenstate of $\sigma_i$ (for $i=x,y$ or $z$) is given by
\begin{equation}
P_i=\Tr{\{\rho (U+\sigma_i)/2\}}
\label{P_i}
\end{equation}
where $\rho$ is the density matrix of the qubit. The probabilities for each of the qubits are computed and shown in Table~\ref{probabilities}.

\begin{table}[H]
\begin{tabular}{l|lll||lll||lll|}
\cline{2-10}
                            & \multicolumn{3}{l||}{Initial}                        & \multicolumn{3}{l||}{After 1st interaction}                      & \multicolumn{3}{l|}{After 2nd interaction}                      \\ \hline
\multicolumn{1}{|l|}{qubit} & \multicolumn{1}{l|}{$\sigma_x$} & \multicolumn{1}{l|}{$\sigma_y$} & $\sigma_z$ & \multicolumn{1}{l|}{$\sigma_x$} & \multicolumn{1}{l|}{$\sigma_y$} & $\sigma_z$ & \multicolumn{1}{l|}{$\sigma_x$} & \multicolumn{1}{l|}{$\sigma_y$} & $\sigma_z$ \\ \hline
\multicolumn{1}{|l|}{A}     & \multicolumn{1}{l|}{1}  & \multicolumn{1}{l|}{0.5}  & 0.5  & \multicolumn{1}{l|}{0.995}  & \multicolumn{1}{l|}{0.505}  & 0.4503  & \multicolumn{1}{l|}{0.9896}  & \multicolumn{1}{l|}{0.5541}  & 0.4558  \\ \hline
\multicolumn{1}{|l|}{B}     & \multicolumn{1}{l|}{0.5}  & \multicolumn{1}{l|}{1}  & 0.5  & \multicolumn{1}{l|}{0.505}  & \multicolumn{1}{l|}{0.995}  & 0.5497  & \multicolumn{1}{l|}{0.505}  & \multicolumn{1}{l|}{0.995}  & 0.5497  \\ \hline
\multicolumn{1}{|l|}{C}     & \multicolumn{1}{l|}{0.5}  & \multicolumn{1}{l|}{0.5}  & 1  & \multicolumn{1}{l|}{0.5}  & \multicolumn{1}{l|}{0.5}  & 1  & \multicolumn{1}{l|}{0.5054}  & \multicolumn{1}{l|}{0.4509}  & 0.9945  \\ \hline \hline
\multicolumn{1}{|l|}{A+B}   & \multicolumn{1}{l|}{1.5}  & \multicolumn{1}{l|}{1.5}  & 1  & \multicolumn{1}{l|}{1.5}  & \multicolumn{1}{l|}{1.5}  & 1  & \multicolumn{1}{l|}{}  & \multicolumn{1}{l|}{}  &   \\ \hline
\multicolumn{1}{|l|}{A+C}   & \multicolumn{1}{l|}{}  & \multicolumn{1}{l|}{}  &   & \multicolumn{1}{l|}{1.495}  & \multicolumn{1}{l|}{1.005}  & 1.4503  & \multicolumn{1}{l|}{1.495}  & \multicolumn{1}{l|}{1.005}  & 1.4503  \\ \hline
\end{tabular}
\caption{The probabilities to measure the positive eigenstate of $\sigma_x$, $\sigma_y$ and $\sigma_z$ for each qubit, in the initial state, after the first interaction (which is between A and B), and after the second interaction (which is between A and C). To ease on the comparison, we wrote on the last two lines the sums of the probabilities of two qubits, which remain constant during a given interaction.}
\label{probabilities}
\end{table}

At the initial state each qubit has a probability of 1 to be measured along the axis it has been prepared (qubit A along $x$ axis, qubit B along $y$ and qubit $C$ along $z$), and 0.5 along any other axis (see Figure~\ref{configuration}). The first interaction is between A and B, hence C is unaffected. After the first interaction A's probability to be in the positive eigenstate of $\sigma_x$ decreases and so does B's probability to be in the positive eigenstate of $\sigma_y$, by the same amount. Each one acquires some of the property of the other: A, B have now a small probability to be in the positive eigenstate of $\sigma_y$ and $\sigma_x$, respectively. There is also a probability exchange to be in the positive eigenstate of $\sigma_z$. 

The second interaction is between A and C, so that B remains unaffected. A and C exchange some probability to be in the positive eigenstate of $\sigma_x$, where A loses what C gains, in the positive eigenstate of $\sigma_y$, where C loses what A gains, and in the positive eigenstate of $\sigma_z$, where C loses what A gains.


\section{Discussion}

We analyzed in this work the policy of adding and tracing out quantum systems for handling multi-interactions of such systems. The conclusion is that a system should be part of a larger system only during the time it interacts. In other words, it may be added before its interaction and traced out after its interaction. We gave an analytic proof of this property and showed a numerical example to demonstrate this principle.

In the context of the FEBERI (free electron bound electron resonant interaction) process of multiple modulation-correlated quantum electron wavefunctions interaction with a TLS \citep{duality, feberi, interrog}, the lesson of this derivation is that the procedure used, of partial tracing of the bound electron state after each interaction, is valid for evaluating the Rabi oscillation evolution of the TLS under a stream of interacting electrons (or its quadratic expansion when starting from ground state \citep{coherent}). In this problem the state of the expired electron is traced out after each electron-TLS dual interaction and the revised TLS state is used for calculating the interaction with the next electron.

Likewise, in the context of interaction of multiple modulation correlated quantum electron wavefunctions with a radiation mode and evolution of bunched-beam superradiance \citep{spont_bin}, the expired electron is traced out after each interaction to provide the updated quantum state of the radiation mode for use in the interaction with the next electron. This provides the evolution of the radiation mode quantum state under the stream of the electrons and the Dicke-kind quadratic growth of the photon number with the number of electrons starting from a vacuum state. These procedures are only limited by the requirement that there is no more than a single electron in the interaction region during its interaction time.

\vspace{6pt} 
\noindent  {\bf Funding:} ``This research was funded by Israeli Science Foundation grant number ISF 2992/24''.

\end{document}